\documentclass[%
 reprint,
 amsmath,amssymb,
 aps,
prx,
]{revtex4-2}

\usepackage{graphicx}% Include figure files
\usepackage{dcolumn}% Align table columns on decimal point
\usepackage{bm}% bold math
\begin{document}

\preprint{APS/123-QED}

\title{Electron-induced nuclear magnetic ordering in n-type semiconductors}% Force line breaks with \\
%\thanks{A footnote to the article title}%

\author{M.~Vladimirova, D.~Scalbert}
 \affiliation{Laboratoire Charles Coulomb, UMR 5221 CNRS/ Universit\'{e}  de Montpellier,
F-34095, Montpellier, France}

\author{M.~S.~Kuznetsova, K.~V.~Kavokin}
\affiliation{Spin Optics Laboratory, St. Petersburg State University, 1 Ul'anovskaya,
Peterhof, St. Petersburg 198504, Russia
}%

\date{\today}% It is always \today, today,
             %  but any date may be explicitly specified

\begin{abstract}
Nuclear magnetism in n-doped semiconductors with positive hyperfine constant is revisited.
Two kinds of nuclear magnetic ordering can be induced by resident electrons in a deeply cooled nuclear spin system.
At positive nuclear spin temperature below a critical value, randomly oriented nuclear spin polarons similar to that predicted by I. Merkulov \cite{Merkulov98} should emerge.
These polarons are oriented randomly  and within each polaron nuclear and electron spins are aligned antiferromagnetically.
At negative nuclear spin temperature below a critical value we predict another type of magnetic ordering -  dynamically induced nuclear ferromagnet.
This is a long-range ferromagnetically ordered state involving both electrons and nuclei.
It can form if electron spin relaxation is dominated by the hyperfine coupling, rather than by the spin-orbit interaction.
Application of the theory to the n-doped GaAs  suggests that the  ferromagnetic order may be reached at experimentally achievable  nuclear spin temperature  $\Theta_\text{N} \approx -0.5$~$\mu$K and lattice temperature $T_\text{L}\approx 5$~K.

%In n-doped semiconductors with positive hyperfine constant, two kinds of magnetically ordered states can be induced by resident electrons in a deeply cooled nuclear spin system.
%
%At positive nuclear spin temperature below a critical value, randomly oriented nuclear spin polarons similar to that predicted by I. Merkulov \cite{Merkulov98} are expected.
%
%At negative nuclear spin temperature below a critical value we predict  the formation of the  long-range ferromagnetic order, if electron spin relaxation is dominated by the hyperfine coupling, rather than by spin orbit interaction.
%
%Application of the theory to n-GaAs with donor concentration  $n_\text{D} \approx 10^{15}$~cm$^{-3}$  suggests that the long-range ferromagnetic order may be reached at realistic nuclear spin temperatures of order of $-2$~$\mu$K and lattice temperature $T=5$~K.
\end{abstract}

%\keywords{Suggested keywords}%Use showkeys class option if keyword
                              %display desired
\maketitle

%\tableofcontents

\section{Introduction}
\label{sec:level1}
%
%Zoology of order states in metals, results in dielectrics, predictions in SCs. \cite{Merkulov1998}

%We consider here the SCs with their particularities (localised electrons).

%- an isolated electron coupled to a large number of nuclear spins under its orbit and resulting ordered states

%- an ensemble of electrons with possible coupling

%Implement the model for the case of

%- an n-doped SC in the insulating regime with varying electron density

%- QDs
%
Magnetism is a very broad subject of condensed matter physics, actively studied owing to its countless applications and its fundamental interest. Current promising research directions include nano-magnetism \cite{Stamps2014}, multi-ferroics \cite{Khomskii2006}, magnetism in graphene \cite{Yazyev2007}, molecular magnetism \cite{Mrozinski2005}, and magnetism in dielectric oxides \cite{Venkatesan2004}, to only cite a few.

\begin{figure}[h!]
  \centering
  \includegraphics[width=8.5 cm]{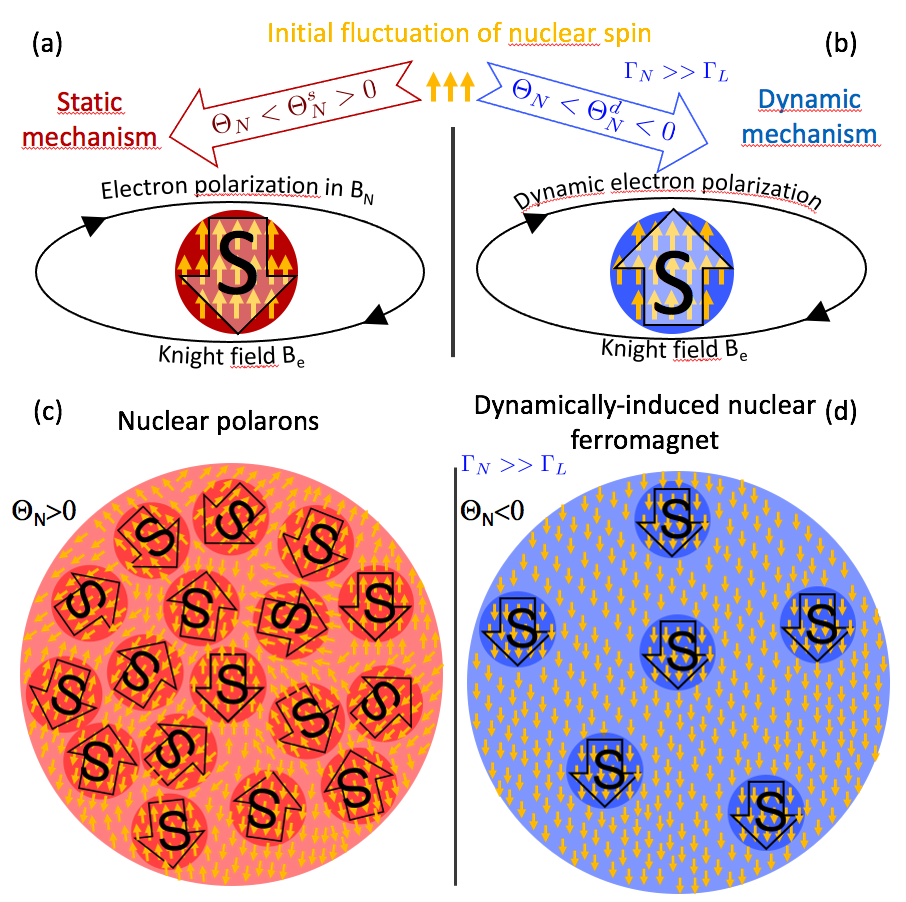}
  \caption{Two kinds of magnetic ordering. \textbf{Left:} At positive nuclear spin temperature local magnetic order (polaron) around each localized electron with spin $S$ may form. The electron spin and the nuclear spins are anti-parallel (a). Polaron spins are oriented randomly, there is no long-range ordering (c).
  \textbf{Right:} At negative nuclear spin temperature electron and nuclear spins are parallel (b). Dynamic polarization of electron spins by the cold NSS may lead to the emergence of the long-range ferromagnetic order (d)  if hyperfine mechanism of electron spin relaxation dominates over  spin-orbit interaction. }\label{figure1}
\end{figure}

Nuclear magnetism is a special case, because interactions between nuclear spins, either dipolar or mediated by hyperfine interaction, are much weaker than electronic spin interactions. For this reason, critical temperatures for nuclear spin ordering in metals or insulators are generally less than 1 $\mu$K \cite{Oja}, except for Van Vleck paramagnets \cite{Ishii2004} and solid He$^3$ \cite{cross1985}, where they are in the mK range.
Nevertheless, since nuclear spin systems (NSSs) offer a reach and original playground in the field of magnetism, they have motivated a large body of research \cite{Oja,Chapellier,Goldman1974,Goldman1977,Abragam,Merkulov98,Merkulov1982,Juntunen2005,Herrmannsdorfer1995,Ishii2004,Roumpos2007,Scalbert2017,Fischer2020}.
Because NSS reaches an internal equilibrium within a time $T_2$, much shorter than the spin-lattice relaxation time $T_2\ll T_1$,  nuclear spins can be cooled down to  temperatures much lower than the lattice temperature \cite{Frohlich,AbragamProctor,AbragamProctor,Kalevich2017}. NSSs also offer a unique opportunity to explore the magnetic phase diagram at negative temperatures \cite{PurcellPound}. In these quite unusual conditions the  thermodynamics tells us that the system tends to maximize its free energy, and antiferromagnetic interactions may lead to a ferromagnetic order \cite{Goldman1977,Oja}.

\begin{figure}[h]
  \centering
  \includegraphics[width=8.5 cm]{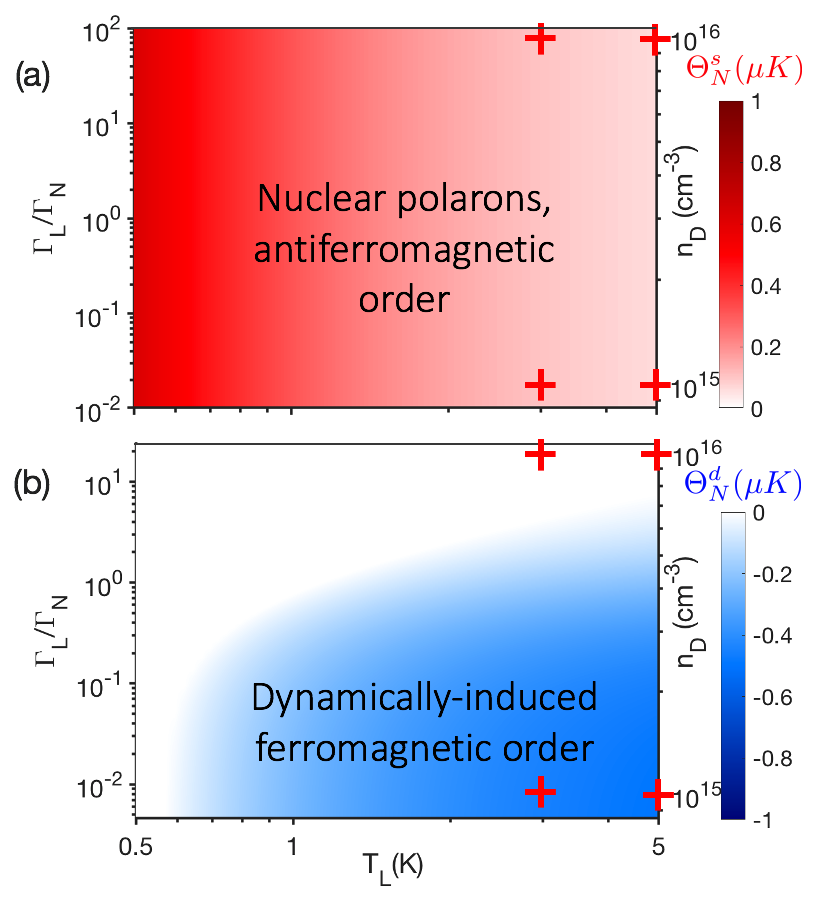}
  \caption{
 Magnetic phase diagram calculated for n-GaAs  NSS cooled down to either positive (a) or negative (b) spin temperature, as a function of the ratio between spin-orbit and hyperfine electron spin relaxation rates $\Gamma_\text{L}/\Gamma_\text{N}$ and lattice temperature $T_\text{L}$.
(a) Randomly oriented nuclear polarons may form in the entire parameter space, the critical temperature $\Theta_\text{N}^\text{s}>0$ is given by Eq.~(\ref{eq:merkulov}).
 (b) Ferromagnetic order emerges in the parameter space given by Eq.~(\ref{eq:thetaNeg}), at $\Theta_\text{N}^\text{d}<0$ given by Eq.~(\ref{eq:critical}).
  Red crosses indicate the points in the parameter space addressed in  Figs.~\ref{figrc} and \ref{fig:spectra}. Right scale shows donor densities corresponding to the values of  $\Gamma_\text{L}/\Gamma_\text{N}$ in GaAs.  %Dashed lines goes through the phase space points where $\Theta_N^{d}\rightarrow 0$.
  %(c)  Color-encoded critical nuclear spin temperature given by Eq.~\ref{eq:critical} for n-GaAs. Below this temperature linearized Eq.~\ref{eq:rate} does not have nontrivial homogeneous static solutions.
%
%  Magnetic phase diagram for the ensemble of donor-bound electrons coupled to the  NSS calculated for  n-GaAs (positive hyperfine constant, $\langle A\rangle=45.6$~$\mu$eV, $I=3/2$) according to Eq.~\ref{eq:critical}.
% Color-encoded critical nuclear spin temperature $\Theta_N^c$ such that linearized Eq.~\ref{eq:rate} does not have nontrivial homogeneous static solutions is shown as a function of the ratio between spin-orbit and hyperfine electron spin relaxation rates $\Gamma_L/\Gamma_N$ and lattice temperature $T_L$.
%  Right scale shows donor densities corresponding to $\Gamma_L/\Gamma_N$ in GaAs (see Section \ref{sec:gaas}). Dashed line separate the regions dominated by static and dynamic polarisation mechanisms.
%  The former results in the formation of the randomly oriented nuclear polarons, while the latter leads to the formation of the carrier-induced ferromagnet.
%  Red crosses (black circles) indicate the points in the parameter space addressed in  Figs.~\ref{figrc}, \ref{fig:spectra} (Fig.~\ref{fig:Sz}), respectively.
}
  \label{fig:isolated}
\end{figure}

Most of the experimental work has been performed in metals, better adapted to demagnetization cooling due to their high thermal conductivity \cite{Oja}. In insulators nuclear spins were first cooled to the milli-Kelvin range by dynamic nuclear polarization using the solid state effect. Final cooling was obtained by adiabatic demagnetization in the rotating frame, to avoid the fast nuclear spin relaxation by paramagnetic impurities which takes place at zero magnetic field \cite{Goldman1974}.
For semiconductors, it was shown theoretically that, similarly to insulators, nuclear magnetic ordering should emerge below a critical
temperature  \cite{Merkulov1982,Merkulov1987}.
%
%The advantage of semiconductors with respect to insulators is double. First, the possibility of efficient coolling of the NSS using the combination of optical pumping via electron spins and adiabatic demagnetisation.
%
%Second, sensible detection of electron and nuclear spin polarization by optical methods \cite{OpticalOrientation,Kalevich2017}.
%
Later, quite different magnetically ordered states have been predicted to form in lightly n-doped semiconductors in the presence of  localized electron states.  The localized states could be either those of shallow donors in n-doped semiconductors in the insulating regime, or weakly strained quantum dots \footnote{In this paper we mainly address  n-GaAs and GaAs/(Al,Ga)As QDs, but these ideas apply to other semiconductors with positive hyperfine constant, such as CdTe or GaN.}.  It was suggested that hyperfine interaction between a localized electron spin and NSS could give rise to the formation of the anti-ferromagnetic ordering in the vicinity of each donor, see Fig.~\ref{figure1}~(a) \cite{Merkulov98,Scalbert2017,Fischer2020}. Such a state was called nuclear spin polaron, in analogy with the magnetic polaron extensively studied (both theoretically and experimentally) in diluted magnetic semiconductors (DMSs) \cite{WolffDMS}. In DMSs the polaron consists of a cloud of spins of magnetic impurities (playing the role of nuclei) ordered under the orbit of a localized electron or hole, and the ordering is induced by the exchange interaction (rather than hyperfine interaction). While the formation of magnetic polarons in DMS has been demonstrated in numerous experiments, the implementation of the polaron in NSS is still awaiting its experimental demonstration.
%
%Previous theoretical works on the nuclear polaron  have assumed the electron spin to be in equilibrium with the crystal lattice, so that electron spin temperature equals the temperature of the lattice  \cite{Merkulov98,Scalbert2017}.
%This assumption corresponds to the situation where electron spin relaxation via spin-orbit interaction (or any other spin lattice interaction mechanism) at rate $\Gamma_{L}$ is far more efficient than the relaxation via hyperfine interaction $\Gamma_{N}$ \cite{Merkulov98,Scalbert2017}.
%Under these conditions randomly oriented nuclear polarons can form in the vicinity of each donor below some critical temperature, see Fig.~\ref{figure1}~(c).
%
In the following the  mechanism underlying the formation of this kind of states  will be referred to as static mechanism, because it involves electron spin relaxation towards thermodynamic equilibrium with the crystal lattice.

In this paper, we extend and amend the existing theory of magnetically ordered states in n-doped semiconductors.
%to the case of an arbitrary relation between hiperfine and spin-orbit relaxation efficiencies.
%This is important because at lattice temperature  $T_L\approx 2$~K in a typical  semiconductor n-doped below metal-to-insulator transition the hyperfine relaxation mechanism can rarely be neglected, and often dominates \cite{Dzhioev2002}.
Our model accounts not only for the electron spin relaxation towards its thermal equilibrium with the lattice, but also  for eventual  dynamic polarization of electrons by the NSS that becomes important when NSS is cooled down to negative temperature \footnote{Here we assume the most widespread case of positive hyperfine coupling constant (relevant to all III-V compounds).}.
We show that if electron spin relaxation via hyperfine interaction dominates over spin-lattice relaxation, long-range  ferromagnetic  order should emerge at negative nuclear spin temperature below a critical value.
The underlying mechanism will be referred to as dynamic mechanism, since it involves dynamic polarization of the electron spins by the NSS.

%This state is characterised by the ferromagnetic ordering of the electron and the nuclear spins, as shown in Fig.~\ref{figure1}~(b), (c). We will refer to the mechanism underlying such ordering as dynamic mechanism, since it involves dynamic polarization of the electron spins  by the NSS.

Taking into account both static and dynamic mechanisms,  we construct the magnetic  phase diagram of the coupled electron-nuclei spin system.
Its implementation for n-GaAs is shown in Fig.~\ref{fig:isolated}.
%Depending on the sign $\Theta_N$, the lattice temperature  $T_L$ and  the  ratio $\Gamma_{L}/\Gamma_{N}$  different kinds of the magnetically ordered states may emerge in a cold NSS.
%of critical temperature for the magnetic ordering $\Theta^c_N $ can be either positive or negative.
 At positive nuclear spin temperature  below  $\Theta_\text{N}^\text{s}>0$, the NSS aligns antiferromagnetically with the electron spin due to static mechanism, so that the ensemble of randomly oriented nuclear polarons emerge.
$\Theta_\text{N}^\text{s}$ decreases when lattice temperature increases, but does not depend on the ratio  $\Gamma_\text{L}/\Gamma_\text{N}$, see Fig.~\ref{fig:isolated}~(a).
At negative nuclear spin temperature  below critical  $\Theta_\text{N}^\text{d}<0$ (Fig.~\ref{fig:isolated}~(b)) a long-range ferromagnetic order builds up in a wide area of the ($\Gamma_\text{L}/\Gamma_\text{N}$, $T_\text{L}$) parameter space. This type of ordering is controlled by the dynamic mechanism, and has been overseen so far.

The paper is organised in seven sections, including appendix.
In the next Section we present a  model describing an ensemble of weakly interacting electron spins localised on shallow donors in a bulk semiconductor, or in QDs, each of them being coupled to the underlying nuclei and an external heat bath (crystal lattice). Rate equations describing this system are derived in Appendix. They allow us to introduce the basic phenomenology of the magnetically ordered states and to identify the  positive feedback loops that govern their formation.
In Section \ref{sec:ensemble} we go beyond the approximation of homogeneous magnetisation, and account for spatial correlations within NSS.
These correlations are critically important, since they determine the nature of the ordered states: the  nuclear polarons ensemble is characterized by zero correlation length, while the ferromagnetic order extends over the entire system.
%
%This  allows us to show that  dynamic polarisation of electrons by the NSS may lead to the long-range ferromagnetic ordering at negative nuclear spin temperature below the critical one.
%While if NSS is cooled down to positive temperature below critical, an ensemble of individual randomly oriented polarons builds up.
%
Two next Sections (\ref{sec:SNSsensing} and \ref{sec:towards}) address the possibility of the experimental detection of the nuclear spin correlations and ordering.
They are followed by concluding remarks.

\section{Phenomenology of the  feedback loop at positive and negative temperatures}
\label{sec:isolated}
%Let us consider spin relaxation of localized electrons interacting with cooled NSS. Two spin relaxation mechanisms should be considered: spin-orbit relaxation at rate $\Gamma_L$  and hyperfine interaction at rate $\Gamma_N$.  and h in the regime of short correlation time $\tau_\text{c}$. This means that  $\tau_\text{c}$ at the given localization center (impurity or quantum dot) is much shorter than the period of its Larmor precession in the Overhauser field created by the rms fluctuation of nuclear spin in the vicinity of this center.
%
%The short correlation time regime assumes that the correlation time   of the electron spin at the given localization center (impurity or quantum dot) is much shorter than the period of its Larmor precession in the Overhauser field created by the rms fluctuation of nuclear spin in the vicinity of this center. The relaxation time of the mean electron spin in the entire ensemble,  , can be much longer than  . This is possible since the electron hopping between centers, as well as the exchange interaction of localized electrons, are nearly spin-conserving. Small spin-orbit corrections to the conduction-band Hamiltonian result in relaxation of the ensemble mean spin with the decay time .

Let us consider spin relaxation of an ensemble of localized electrons interacting with NSS cooled down to temperature $\Theta_\text{N}<T_\text{L}$. Each electron spin interacts with $N$ nuclei.
The electron spin correlation time  $\tau_\text{c}$ is supposed to be short. This means that $\tau_\text{c}$ at a given localization center (impurity or QD) is much shorter than the period of the electron spin precession in the Overhauser field $B_\text{N}$ created by the random fluctuations of nuclear spin in the vicinity of this center. In this case the relaxation time of the entire electron spin ensemble is longer than $\tau_\text{c}$, because electron hopping between centers, as well as the exchange interaction between localized electrons, are nearly spin-conserving. Small spin-orbit corrections to the conduction-band Hamiltonian lead to the relaxation of the ensemble mean spin at rate $\Gamma_\text{L}<<1/\tau_\text{c}$.
The regime of short correlation time is relevant in the majority of experiments on the electron-nuclear spin dynamics in bulk semiconductors and nanostructures, with the exception of single quantum dots \cite{KavokinSST09,Dzhioev2002,Belykh2017}.

Due to some fluctuation, the average electron spin $\langle \vec{S} \rangle$ (supposed to be homogeneous in space) may differ from zero $\langle \vec{S} \rangle=\langle {S}_z \rangle \vec{e_z}$. Then, the Knight field $B_e$ created by non-zero electron spin gives rise to the nuclear spin polarization and thus the average nuclear spin in the same direction.
%$\langle {J}_z \rangle=-I(I+1)\beta_N \langle A \rangle \langle {S}_z \rangle/3N$.
The dynamics of this ensemble can be described by the following rate equation:
\begin{equation}
\langle \dot{{S}_z} \rangle =-\Gamma_\text{S}(\langle {S}_z \rangle-{S_\text{T}})+ \Gamma_\text{N}\frac{\langle {J}_z \rangle }{2 \langle {J}_{\perp}^2 \rangle }
\left ( 1-\frac{\langle {S}_z\rangle {S_\text{T}}}{S^2}
\right)
\label{eq:rate}
\end{equation}
Here  $\langle {J}_{\perp}^2 \rangle$ is the mean squared transverse (perpendicular to the Knight field) fluctuation of the total nuclear spin interacting with the electron, $S=1/2$ is the electron spin value, $S_\text{T}$ is the equilibrium value of the electron spin in the presence of the spin-polarized nuclei at a given lattice temperature $T_\text{L}$.
Derivation of Eq.~(\ref{eq:rate})  from the  basic laws of quantum mechanics is provided in Sec.~\ref{Sec:Eq1}.

In this work  we limit our considerations to the case of weak polarization of electron and nuclear spins, which remains relevant until  collective electron-nuclei spin states are not  formed.
In this approximation $\langle {S}_z\rangle {S_\text{T}}\approx0$ and $2 \langle {J}_{\perp}^2 \rangle \approx 4 I(I+1)/3\equiv Q$.
Thus,  Eq.~(\ref{eq:rate}) reduces to:
\begin{equation}
\langle \dot{{S}_z} \rangle =-\Gamma_\text{S}(\langle {S}_z \rangle-{S_\text{T}})+ \Gamma_\text{N}\frac{\langle {J}_z \rangle }{Q }
\label{eq:rateLin}
\end{equation}
 Its first term on the right-hand side accounts for the relaxation of the electron mean spin towards its value at thermal equilibrium with the lattice, $S_\text{T}$, at the rate $\Gamma_\text{S}=\Gamma_\text{L}+\Gamma_\text{N}$.
The second term is related to electron-nuclei spin flips.
This term was not considered in the nuclear magnetism models developed previously.
It allows for the dynamic polarization of the electron by the cold nuclei and is responsible for out-of-equilibrium electron spin polarization.
Assuming that the wavefunction of the localised electron has a spherically symmetric  exponential form characterised by the Bohr radius $a_\text{B}$ we can write the average nuclear spin projection on the Knight field as
\begin{equation}
\langle J_z \rangle =-\frac{ I (I+1)}{3N}\langle {S}_z \rangle
\langle {A} \rangle \beta_\text{N}.
\label{eq:JzN}
\end{equation}
Here $I$ is the nuclear spin value (assumed to be identical for all nuclear species in the crystal),  $\beta_\text{N}$ is the inverse nuclear spin temperature expressed in energy units,  $\beta_\text{N}=1/k_\text{B} \Theta_\text{N}$, $k_\text{B}$ is the Boltzmann constant, $\langle A \rangle=\sum_l{\mathcal{A}_lA_{l}}$    is the hyperfine interaction constant averaged over all nuclear species in the crystal, $A_{l}$ and $\mathcal{A}_l$ are the hyperfine constant and the abundance of $l$-th isotope, respectively, $N=27\pi a_\text{B}^3/8v_0$ is the number of nuclei under the donor orbit, $v_0$ is the volume of the crystal elementary cell.
Within the same approximation the electron spin polarization at equilibrium, $S_\text{T}$, reads:
\begin{equation}
S_\text{T} =-\frac{\langle {A} \rangle \beta_\text{L}}{4} \kappa \langle {J}_z \rangle,
\label{eq:ST0}
\end{equation}
 where $\beta_\text{L}=1/k_\text{B} T_\text{L}$ is the inverse lattice temperature expressed in energy units, $\kappa=27 n_0/64$ and $n_0$ is the number of atoms in the crystal elementary cell.
 %
 %Thus,  Eq.~(\ref{eq:rate}) reduces to:
%\begin{equation}
%\langle \dot{{S}_z} \rangle =-\Gamma_S(\langle {S}_z \rangle-{S_T})+ \Gamma_N\frac{\langle {J}_z \rangle }{Q }
%\label{eq:rateLin}
%\end{equation}
%
%The first term on the right-hand side of Eq.~(\ref{eq:rateLin}) accounts for the relaxation of the electron mean spin towards its value at thermal equilibrium with the lattice, $S_T$, at the rate $\Gamma_S=\Gamma_L+\Gamma_N$.
%
%The second term is related to electron-nuclei spin flips.
%
%This term was not considered in the nuclear magnetism models developed previously.
%
%It allows for the dynamic polarization of the electron by the cold nuclei and is responsible for out-of-equilibrium electron spin polarization.
%

%
%

Equation~(\ref{eq:rateLin}), with $\langle J_z \rangle$ and $S_\text{T}$ given by Eqs.~(\ref{eq:JzN}) and (\ref{eq:ST0}), may have non-trivial static solutions.
The static solution of Eq.~(\ref{eq:rateLin}) $\langle \vec{J} \rangle=\langle \vec{S}\rangle=0$ becomes unstable at some critical value of the  nuclear spin temperature
\begin{equation}
k_\text{B}\Theta_\text{N}^\text{c}=  \frac{Q \kappa  \langle A \rangle^2 \beta_\text{L}}{16 N}-\frac{ \langle A \rangle}{4 N}\frac{\Gamma_\text{N}}{\Gamma_\text{S}},
\label{eq:critical}
\end{equation}
%
%Critical temperatures given by Eq.~\ref{eq:critical} for GaAs  are shown in Fig.~\ref{fig:isolated}~(c).
%
In the case $ \langle A \rangle<0$, $\Theta_\text{N}^\text{c}$ is always positive. The static and dynamic mechanisms are both acting in concert to achieve a collective nuclear spin state. Whereas if $ \langle A \rangle>0$,  $\Theta_\text{N}^\text{c}$ can be either positive or negative depending on both lattice temperature and the ratio $\Gamma_\text{N}/\Gamma_\text{L}$.

In the limit where dynamic polarisation of electrons by the cold NSS can be neglected (the second term  in Eq.~(\ref{eq:rateLin}) is  close to zero if $\Gamma_\text{N} \ll 1$) Eq.~(\ref{eq:critical}) yields the positive value of the critical temperature
\begin{equation}
k_\text{B}\Theta_\text{N}^\text{s}=\frac{Q \kappa \langle A \rangle^2 \beta_\text{L}}{16 N}
\label{eq:merkulov}
\end{equation}
corresponding to the formation of the polaron state  via static mechanism only, as first described by I.~Merkulov \cite{Merkulov98}.
%
%Such state can form in an individual quantum dot coupled to  electron gas, if the spin relaxation by cotunneling is the fastest mechanism. In this case, dynamic polarization is suppressed, and the polaron formation mechanism via thermal polarization of the electron spin in the Overhauser field is the only one to be considered. \cite{Merkulov98}.
%
%Note also, that in n-doped semiconductors characterized by negative hyperfine constant $\langle A \rangle<0$ (e.g. CdTe) the critical temperature is always positive, for any values of electron spin relaxation rates.

The formation mechanism of the ordered state at positive nuclear spin temperature is similar to that of  the polaron predicted by I.~Merkulov, or the magnetic polaron in DMSs \cite{WolffDMS}.
It can be understood in terms of effective fields, the nuclear (Overhauser) field acting on the electron spin and the electron (Knight) field $B_e$ acting on the nuclei.
Let us suppose that the electron spin gets polarized to its thermal equilibrium value in a fluctuation of the nuclear field.
%This implicitly means that the electron spin is well coupled to the lattice.
The Knight field created by such polarized electron acts on the nuclear spins, enhancing the initial fluctuation. This closes the feedback loop and, if the gain is larger than one, the initial fluctuation will grow until a nuclear polaron is formed. If, like in GaAs, the hyperfine coupling constant $\langle A \rangle$ is positive, the electron polarization is anti-parallel to the nuclear spins. We would like to point out that directions of net spins of different static polarons need not be correlated, because the electron spin at each site tends to relax to its equilibrium value in the local Overhauser field.

However, the formation of randomly oriented polarons cannot be consistently described by Eq.~(\ref{eq:critical}) obtained assuming homogeneous average spin polarization.
Thus, one should go beyond this approximation and consider spatial correlations between nuclear spins at different electron sites. This is done in the next Section, where we show that at $\Theta_\text{N}>0$ the magnetic ordering occurs in the form of randomly oriented nuclear polarons even in presence of dynamic polarization (i.e. at nonzero $\Gamma_\text{N}$). The instability arises at $\Theta_\text{N}$ equal to $\Theta_\text{N}^\text{s}>0$  given by Eq.~(\ref{eq:merkulov}), see Fig. \ref{fig:isolated}~(a).

The mechanism responsible for magnetic ordering at negative spin temperature is efficient if the electron spin is loosely coupled to the lattice, so that the static mechanism of electron polarisation is overcome by the dynamic one in Eq.~(\ref{eq:rate}). In this case the electron spin polarization is always parallel to that of nuclear spins, in contrast with the static polaron case.  At negative $\Theta_\text{N}$ this provides the positive feedback loop (Fig.~\ref{figure1}~(b, d)).  The ferromagnetic alignment  of the NSS and electrons builds up below the critical temperature given by Eq.~(\ref{eq:critical}), provided that the ratio $\Gamma_\text{N}/\Gamma_\text{L}$ is big enough. Thus, the conditions for the ferromagnetic ordering read:
\begin{align}
k_\text{B}\Theta_\text{N}^\text{d}= \frac{Q \kappa \langle A \rangle^2 \beta_\text{L}}{16 N}-\frac{ \langle A \rangle}{4 N}\frac{\Gamma_\text{N}}{\Gamma\text{S}}, \nonumber \\
\frac{\Gamma_\text{L}}{\Gamma_\text{N}}<\frac{4}{Q \kappa \langle A \rangle \beta_\text{L}}-1,
\label{eq:thetaNeg}
\end{align}
 In contrast to the static mechanism, the dynamic mechanism involves the onset of the net spin polarization in the ensemble of electrons, since in the regime of short correlation time the non-equilibrium electron spin is spread over a large number of localization centers.
We will show in the next Section that this kind of  magnetic order expands over the entire system, so that the resulting long-range state can be qualified  as a carrier-induced nuclear ferromagnet (Fig.~\ref{fig:isolated}~(b)).

The above considerations allow us to make some predictions about magnetic ordering as a function of the nuclear spin temperature $\Theta_\text{N}$, lattice temperature $T_\text{L}$ and electron spin correlation time $\tau_\text{c}$ that governs the ratio $\Gamma_\text{N}/\Gamma_\text{L}$.
However, in order to quantify the spatial extension of these states (which, as it was anticipated depends on the sign of the nuclear spin temperature) one should address the spatial dependence of the NSS susceptibility.  \cite{KavokinSST09,Henn2013}.
This is the subject of the next Section.

%\begin{figure}
 % \centering
%  \includegraphics[width=7.5 cm]{feedback_loop.pdf}
%  \caption{Illustration of the two feedback loops which tend to amplify an initial nuclear spin fluctuation at negative and positive temperatures.}\label{feedbackloop}
%\end{figure}

%\section{Magnetic states formed by an isolated electron coupled to a large number of nuclear spins}
%\subsection{Rate equation for an isolated donor}
%\subsection{Evidence of two polarons}
%\subsection{Implementation to QDs and very low density n-type bulk}

\section{Spatial dependence of the electron-induced nuclear magnetization: randomly oriented polarons versus nuclear ferromagnet}
\label{sec:ensemble}
%
%\subsection{Theory}
%\label{sec:theory}
The mean spin of electrons as function of time and spatial coordinates (on the spatial scale much greater than average distance between donors) obeys the continuity equation:
\begin{multline}
\langle \dot{{S_z}}(\vec{R_n}) \rangle=
-\Gamma_\text{S}(\langle {S_z}(\vec{R_n}) \rangle -  S_\text{T}(\vec{R_n}))+
\frac{\Gamma_\text{N} \langle {J_z}(\vec{R_n}) \rangle}{Q} +\\
 \mathrm{div}\left[ D_\text{s} \nabla(\langle {S_z}(\vec{R_n})\rangle- S_\text{T}(\vec{R_n}) ) \right],
 \label{eq:continuty}
\end{multline}
 were $\vec{R_i}$ defines the n-th donor coordinate in space, while $\langle {S_z}(\vec{R_n}) \rangle$, $ {S_\text{T}}(\vec{R_n}) $  stand for mean and equilibrium values of the electron  spin projections at n-th donor, respectively  and $\langle {J_z}(\vec{R_n})\rangle $ is the mean nuclear spin at n-th donor, see Eqs.~(\ref{eq:JzN})-(\ref{eq:ST0}).
 Eq.~(\ref{eq:continuty}) is analogous to Eq.~(\ref{eq:rateLin}), but includes an additional term. It accounts for the electron spin diffusion between donors, which is characterized by the diffusion constant $D_\text{s}$.
 %

%
%The first term on the right side describes electron spin relaxation towards its thermal equilibrium state  $S_T(t,\vec{R})=-\langle A\rangle \beta_L\langle {\vec{J}}(t, \vec{R}) \rangle /4$ in the local Overhauser field.
%The first term on the right side describes electron spin relaxation towards its thermal equilibrium state.
%
%Here $\langle A\rangle=\sum_l{\mathcal{A}_l A_{l}}$ is the hyperfine constant averaged over all nuclear isotops.
%
%The second term accounts for the electron spin diffusion, characterized by the diffusion constant $D_\text{s}$.
%
%The last term describes the dynamic polarization of electron spin by the NSS.
%
%As long as the  ordered states are not formed we may use  the relation $\langle I_z^2 \rangle =I(I+1)/3$, so that $\kappa=1/Q$, and neglect the  vanishing contribution proportional to  $\langle \vec{S}\rangle S_T$.
%It is given by the sum of spin-orbit relaxation rate $\Gamma_L=3D_\text{s}/L_{SO}^2$ and hyperfine relaxation rate $\Gamma_N=2\Omega_N^2\tau_\text{c}/3$, where $\Omega_N$ is the electron spin precession frequency in the local Overhauser field. Finally, $D_\text{s}= \frac{n_\text{D}^{-2/3}}{3\tau_\text{c} }$ is the electron spin diffusion constant, that depends on the donor density $n_\text{D}$.
%
In Fourier components Eq.~(\ref{eq:continuty}) reads:
\begin{equation}
i\omega \langle {S_z} \rangle_{\omega,\vec{k}}=
-\Gamma_\text{S} {\Upsilon}  + \Gamma_\text{N} \frac{\langle {J_z} \rangle_{\omega,\vec{k}}}{Q} - D_\text{s} k^2 {\Upsilon},
\label{eq:fourierCont}
\end{equation}
where, $\vec{\Upsilon}=\langle \vec{S} \rangle_{\omega,\vec{k}}+
\kappa \langle A\rangle \beta_\text{L}\langle \vec{J} \rangle_{\omega,\vec{k}} / 4$ and
\begin{align}
\langle {J_z} \rangle _{\omega, \vec{k}} =
N_\text{D}^{-1}\sum_n \langle {J_z}(\vec{R}_n) \rangle_\omega \exp(i \vec{k}\cdot\vec{R}_i),     \\
\langle {S_z} \rangle _{\omega, \vec{k}} =
N_\text{D}^{-1}\sum_n \langle {S_z}(\vec{R}_n)  \rangle_\omega \exp(i \vec{k}\cdot\vec{R}_n),
\end{align}
where $N_\text{D}$
%=\Sigma n_\text{D}$
is the number of donors in the sample, and we assume that $k<<n_\text{D}^{1/3}$, $n_\text{D}$ being the concentration of the donors.
Since the number of nuclei interacting with one electron, $N$, is large, and electron spin dynamics is much faster than that of nuclei, the effect of electron-nuclear interaction on the nuclear spin susceptibility $\chi(\omega, \vec{k}$) can be considered in the mean-field approximation. This way, the Fourier components of nuclear and electron spin at the n-th donor are related via
\begin{equation}
\langle {{J_z}(\vec{R}_n)} \rangle _\omega = \frac{\chi(\omega) }{N n_\text{D}}
\left(  b_1 \exp(i \vec{k}\cdot\vec{R}_n) + \overline{b}_e \langle {S_z}(\vec{R}_n)\rangle _\omega \right),
\label{eq:Jomk}
\end{equation}
where $\overline{b}_e =-\langle A \rangle / (N \hbar \langle \gamma_\text{N} \rangle)$ is the Knight field at saturation, $b_1$ is an arbitrary oscillating field parallel to it, $\gamma_\text{N}$ is the nuclear gyromagnetic ratio averaged over the nuclear species in the crystal, and $\hbar$ is the reduced Plank constant.

Since we are interested in the “nuclear” scale of frequencies, the condition $\omega/\Gamma_\text{S} << 1$ is always fulfilled, and we can put the left-hand side of Eq.~(\ref{eq:fourierCont})  equal to zero.  Then, from Eqs.~(\ref{eq:fourierCont}) and (\ref{eq:Jomk}) we obtain:
\begin{gather}
\langle {S_z} \rangle_{\omega,\vec{k}}=\langle {J_z} \rangle_{\omega,\vec{k}}
\zeta(\vec{k}) \\
\langle {J_z} \rangle_{\omega,\vec{k}}=\frac{1}{N n_\text{D}}\frac{\chi(\omega)}
{1-\zeta(\vec{k})\overline{b}_e \chi(\omega)/(N n_\text{D})}{b}_1,
\label{eq:SJomk}
\end{gather}
where
\begin{equation}
\zeta(\vec{k})=-\frac{\langle A\rangle \kappa \beta_\text{L}}{4} +
\frac{\Gamma_\text{N}}{Q  (\Gamma_\text{S}+D_\text{s}  {|\vec{k}|}^2)}.
\label{eq:zeta}
\end{equation}

Eq.~(\ref{eq:SJomk}) allows one to calculate the $\vec{k}$-dependence of the total fluctuation power $\langle J_z^2 \rangle_{0,\vec{k}}$, as well as the total static susceptibility of the nuclear spin $\chi_{0,\vec{k}}$:
\begin{equation}
\langle J_z^2 \rangle_{0,\vec{k}}=\frac{Q/4}
{1+\zeta(\vec{k})\langle A \rangle \beta_\text{N} Q/4N}
\label{eq:fluct}
\end{equation}
\begin{equation}
{\chi}_{0,\vec{k}}=N n_\text{D}\hbar \langle \gamma_\text{N} \rangle \langle J_z^2 \rangle_{0,\vec{k}}
\label{eq:susc}
\end{equation}

%\begin{figure}
 % \centering
 % \includegraphics[width=8.5 cm]{fig_ensemble.png}
%  \caption{Phase diagram of nuclear spin ordering model in n-GaAs (positive hyperfine constant, $\langle A\rangle=45.6$~$\mu$eV, $J=3/2$).
%  Color-encoded critical nuclear spin temperature $\Theta_N^c$ for the formation  of carrier-induced nuclear ferromagnet ($\Theta_N^c>0$) (a)  and randomly oriented nuclear polarons ($\Theta_N^c>0$)(b). The parameter space is defined by the ratio between spin-orbit and hyperfine electron spin relaxation rates $\Gamma_L/\Gamma_N$ and lattice temperature $T_L$. Red crosses indicate the points in the parameter space corresponding to Figs.~\ref{figrc} and \ref{fig:spectra}. }
%  \label{fig:ensemble}
%\end{figure}

The divergence of the susceptibility is a signature of the collective state formation. The spatial correlation function of the nuclear spin fluctuations is given by the Fourier image $\mathcal{F}(r)$ of Eq.~(\ref{eq:fluct}). It contains all the information on the spatial ordering of the nuclear spin, including its correlation length $r_\text{c}$.
\begin{equation}
\mathcal{F}(r)=\frac{\delta(r)}{1-\beta_\text{N}/\beta_\text{N}^s}-
\frac{\langle A\rangle\Gamma_\text{N}}
{4 N D_\text{s}}\times\frac{\beta_\text{N}}
{(1-\beta_\text{N}/\beta_\text{N}^\text{s})^2}\times
\frac{e^{-r/r_\text{c}}}{4\pi r},
\label{eq:fourier}
\end{equation}
with
\begin{equation}
r_\text{s}=\sqrt{\frac{D_\text{s}}{\Gamma_\text{S}} \times \frac{1-\beta_\text{N}/\beta_\text{N}^\text{s}}{1-\beta_\text{N}/\beta_\text{N}^\text{d}}}
\label{eq:rc}
\end{equation}
where we defined the inverse critical temperatures as $\beta_\text{N}^\text{d}=1/(k_\text{B} \Theta_\text{N}^\text{d})$ and  $\beta_\text{N}^\text{s}=1/(k_\text{B} \Theta_\text{N}^\text{s}$).
The first term in Eq.~(\ref{eq:fourier}) corresponds to the absence of any  correlations between nuclear spins situated under the orbits of two different donors, while the second gives the contribution of carrier-induced magnetic ordering on the scale of $r_\text{c}$.
\begin{figure}
  \centering
  \includegraphics[width=8.5 cm]{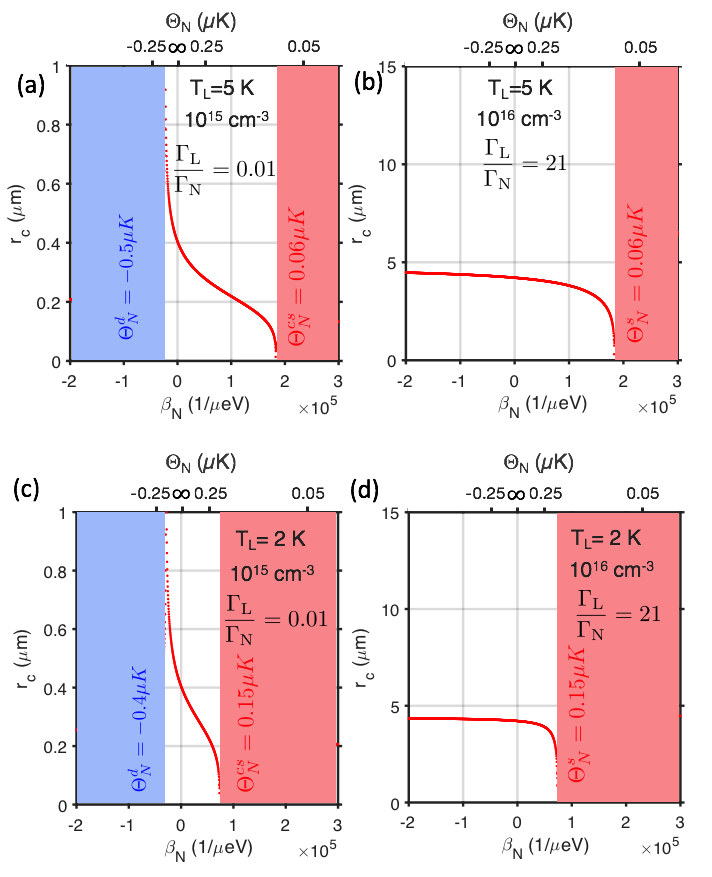}
  \caption{Correlation length $r_\text{c}$ calculated using Eq.~(\ref{eq:rc}) as a function of nuclear spin temperature $\Theta_\text{B}$ (upper scale) or inverse nuclear spin temperature $\beta_\text{N}$ (lower scale). Two values of the lattice temperature $T_\text{L}=5$~K (a, b), $T_\text{L}=2$~K (c,d) and two different donor densities $n_\text{D}= 10^{15}$~cm$^{-3}$ (a, c), $n_\text{D}=10^{16}$~cm$^{-3}$ (b, d) are shown.}
  \label{figrc}
\end{figure}

\begin{figure}
  \centering
  \includegraphics[width=8.5 cm]{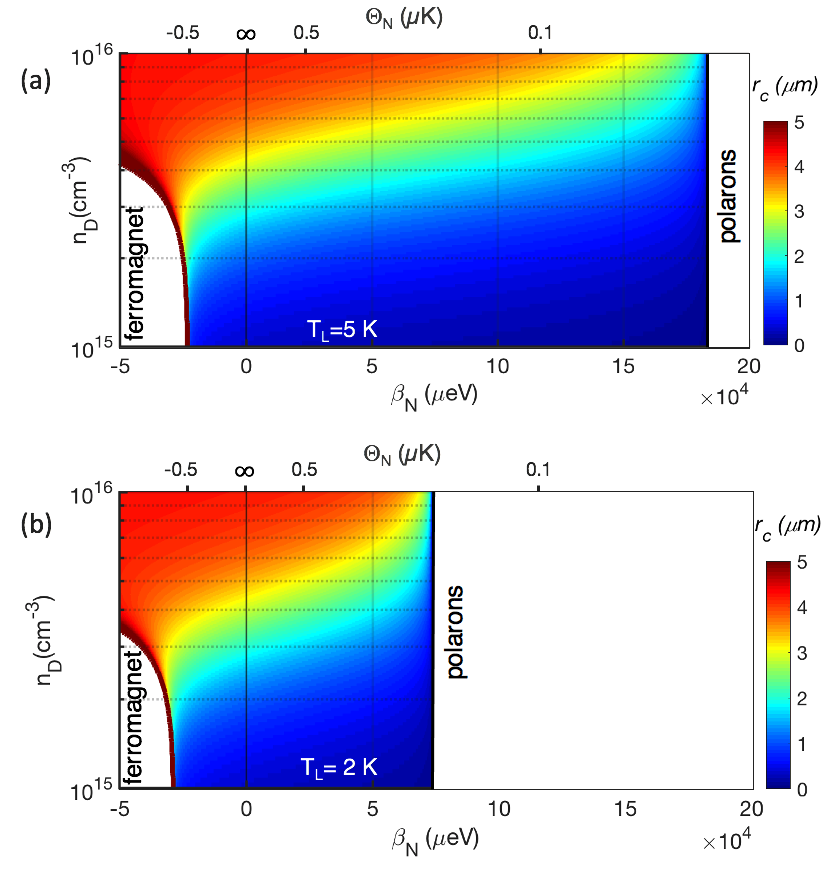}
  \caption{Color-encoded correlation length $r_\text{c}$ calculated  for n-GaAs using Eq.~(\ref{eq:rc}) at $T_\text{L}=5$~K (a) and $T_\text{L}=2$~K (b) as a function of donor density and nuclear spin temperature.}
  \label{fig:rcnD}
\end{figure}

It is easy to see that both the correlation length $r_\text{c}$, and the static susceptibility to uniform magnetic fields, $\chi_{0,\vec{k}=0}$, diverge  at negative critical temperature  $\Theta_\text{N}^\text{d}$, where ferromagnetic ordering is expected due to dynamic feedback mechanism.
Thus, because at  $\Theta_\text{N}^\text{d}$ the correlation length $r_\text{c}\rightarrow \infty$, the dynamic mechanism  leads to the formation of a long-range ferromagnetic  order, as sketched in Fig. \ref{figure1}~(d).
%
%In the limit of $\Gamma_L/\Gamma_N<<1$ above the critical temperature the correlation radius decreases, and reaches zero at temperatures given by Eq.~\ref{eq:merkulov}.
%
%There is only a specific range of the $\Gamma_L<<\Gamma_N$ ratio values where such magnetic order can build up.

At $\Theta_\text{N}^\text{s}>0$ given by Eq.~(\ref{eq:merkulov}), where static mechanism dominates over the dynamic one, and the electron spin aligns anti-ferromagnetically with nuclei,  the correlation function $\mathcal{F}(r)$ diverges, while the correlation length  goes to zero, $r_\text{c}\rightarrow 0$. This means that local nuclear spin fluctuations grow in amplitude, remaining spatially uncorrelated. Eventually, these fluctuations develop into magnetic polarons.
Thus, static mechanism of electron-nuclei interaction leads to the formation of the individual polaron states with random spin orientation sketched in Fig.~\ref{figure1}~(d).
Remarkably, the values of corresponding critical temperature $\Theta_\text{N}^\text{s}$ given by Eq.~(\ref{eq:merkulov}) are those that one would obtain by simply neglecting the dynamic polarization term in Eq.~(\ref{eq:rateLin}) \cite{Merkulov98,Scalbert2017,Fischer2020}.
This means that dynamic polarization does not alter the formation of the individual randomly oriented polarons. The reason for this is that in the ensemble of randomly oriented polarons the net nuclear spin is zero, and no directional transfer of angular momentum into the electron spin system occurs.

The values of positive and negative critical temperature obtained in this framework are color-encoded in Figs.~\ref{fig:isolated}~(a,b)  as a function of lattice temperature and $\Gamma_\text{L}/\Gamma_\text{N}$. These magnetic phase diagrams represent the main result of this paper.

%To obtain analytical expressions for the magnetic phase diagrams shown in Figs. \ref{fig:isolated}~(a, b) we neglected the nonlinear conribution in the dynamic polarization term of Eqs.~(\ref{eq:rate}) and (\ref{eq:continuty}).
%To get an insight in the role of these nonlinear terms (only important at negative spin temperatures) the full nonlinear equation must be solved numerically. Such a numerical analysis shows that the linear approximation works fairly well when $\Theta_{N}^{cs}$ predicted by the linear model is not too low.
%
%At very low spin temperatures of the NSS, the nonlinear contribution strengthens the dynamic mechanism.
%
%As a result, the phase diagram in Fig.~\ref{fig:isolated}~(b) is slightly modified:  the absolute values of the critical temperature are higher that that predicted by the linear model;
%
%We report in  Appendix (Sec.~\ref{Sec:nonlin}) two representative examples of the numerical solution of  Eq.~(\ref{eq:rate}), but the detailed  study of the system in the region of the lowest spin temperatures remains below the scope of this paper.

Using the  parameters of the NSS in n-GaAs (summarized in  Appendix \ref{sec:gaas}) we  represent in Fig.~\ref{figrc} the critical length $r_\text{c}$ calculated for two values of the lattice temperature and two donor densities (the corresponding points of the parameter space are indicated by red crosses in Fig.~\ref{fig:isolated}).
%
%The correlation length is represented as a function of both inverse nuclear spin temperature $\beta_N$ (bottom scale) and nuclear spin temperature itself (upper scale).
%
One can see that the correlation length varies monotonously as a function of the inverse nuclear spin temperature: from infinity at $\Theta_\text{N}^\text{d}<0$  where ferromagnetic order is expected, to zero at $\Theta_\text{N}^\text{s}>0$ where nuclear polarons emerge.
If the parameters of the system  are such that the ferromagnetic order can never emerge (white area in Fig.~\ref{fig:isolated}~(b), the correlation length does not diverge at negative temperature. This is illustrated in Fig.~\ref{figrc}~(d)).

In the limit of high nuclear spin temperature ($\beta_\text{N}\rightarrow0$) the correlation length is given by  $r_\text{c}^{\infty}=\sqrt{ D_\text{s}/\Gamma_\text{S}}$.
It is governed by the interplay between electron spin flip and diffusion efficiency. $r_\text{c}^{\infty}$  depends on the donor concentration (see Fig.~\ref{fig:rcnD}) and can be interpreted as a spin diffusion length.

Note also, that regardless the type of magnetic ordering, the correlation length varies strongly in the vicinity of the ordering transition.
This is illustrated in Fig.~\ref{fig:rcnD}. It shows the correlation length as function of nuclear spin temperature and donor density at $T_\text{L}=5$~K (a) and  $T_\text{L}=2$~K (b). The parameters of the calculation are given in Appendix \ref{sec:gaas} ).
On can see that the variation of the critical length with nuclear spin temperature can reach several  micrometers.
This suggests that  even above the critical temperature these correlations may be detected, {\it e.g} via spin noise spectroscopy.
This possibility is analysed in the next Session.

\section{Sensing nuclear spin correlations and ordering by spin noise spectroscopy}
\label{sec:SNSsensing}
One of the promising methods that can be used to evidence electron-induced nuclear spin ordering is the electron spin-noise spectroscopy (SNS) \cite{Romer2007,Hubner2014,Cronenberger2016,Cronenberger2019}.
 SNS is based on the fluctuation-dissipation theorem, which states that it is possible to detect resonances of linear susceptibility  by ``listening`` to a noise of the medium in its equilibrium state.
It allows one to probe electron spin fluctuations non-perturbatively using absorption-free Faraday rotation.
The Faraday rotation noise spectrum features a peak at the magnetic resonance frequency $\nu_\text{L}$ corresponding to precession of spontaneous fluctuations of the spin ensemble at the Larmor frequency.
The latter is given by the total magnetic field acting on the electron, that is a sum of the  external field and  the Overhauser field $B_\text{N}$ \cite{Ryzhov2015}.
Thus, one can expect that the  formation of the ordered state at $B=0$ will be accompanied by the shift of the electron spin-noise spectrum peak from zero to $\gamma_eB_\text{N}$, where $\gamma_e$ is the electron gyromagnetic ratio.

Even above the critical temperature, the correlations induced by the electrons in the  deeply cooled nuclear spin system can be detected via SNS. One could detect variations of the correlation length in the electron spin fluctuations in the vicinity of the critical temperature by the recently developed spatiotemporal spin noise spectroscopy \cite{Cronenberger2019}. Another possibility would be to detect directly the nuclear spin noise \cite{Berski2015}. Below we study how the correlations in the NSS affect the shape of the electron spin noise spectrum.

The spectral power density $\langle S_z^2(\omega)\rangle$ of electron spin fluctuations can be expressed in terms of the total nuclear spin fluctuation power$\langle J_z^2 \rangle_{0,\vec{k}}$ given by Eq.~(\ref{eq:fluct}), normalized by the square of the total spin value $\overline{\langle J_z^2 \rangle_{0,\vec{k}}}=3 \langle J^2 \rangle_{0,\vec{k}}/I(I+1)$:
\begin{multline}
\langle S_z^2(\omega)\rangle=
\left[
\left( \frac{\Omega_\text{N}^2}{3\pi^2 n_\text{D}} \int_0^{k'}
\frac{D_\text{s} k^4 \times \overline{\langle J_z^2 \rangle_{0,\vec{k}}} }{(D_\text{s} k^2)^2+\omega^2} d k\right)^2+ \right.\\
\left.+\left( \omega-\frac{\Omega_\text{N}^2}{3\pi^2 n_\text{D}} \int_0^{k'}
\frac{\omega \times  \overline{\langle J_z^2 \rangle_{0,\vec{k}}} }{(D_\text{s}k^2)^2+\omega^2}dk\right)^2 \right]^{-1},
\label{eq:SNS}
\end{multline}
where $k'=2/3 \pi^2 {n_\text{D}^{-1/3}\Gamma_\text{S}}/\Gamma_\text{N}$.
%2/3 (\pi^2 \Gamma_\text{s}  / \Gamma_N)
%
The derivation of this expression is given in Appendix \ref{Sec:SNS}.

\begin{figure}
  \centering
  \includegraphics[width=8.5 cm]{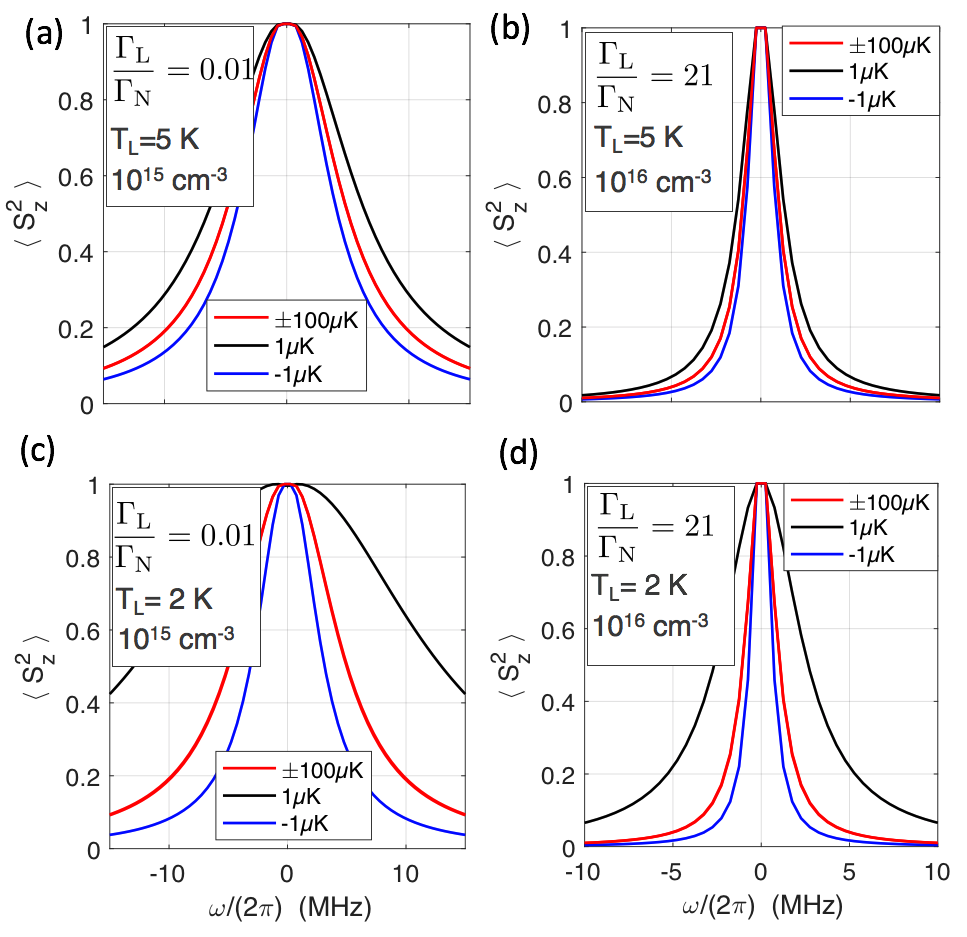}
  \caption{Electron spin noise spectra calculated using Eq.~\ref{eq:SNS}  at four different values of the nuclear spin temperature: $\Theta_\text{N}=\pm 100$~$\mu$K (these two spectra are almost identical) and $\Theta_\text{N}=\pm 1$~$\mu$K. Two values of the lattice temperature $T_\text{L}=5$~K (a,b), $T_\text{L}=2$~K (c,d) and two different donor densities $n_\text{D}= 10^{15}$~cm$^{-3}$ (a, c), $n_\text{D}=10^{16}$~cm$^{-3}$ (b, d) are shown.}
  \label{fig:spectra}
\end{figure}
Fig.~\ref{fig:spectra} shows electron spin noise spectra calculated in n-GaAs  for the same combinations of doping and lattice temperature (shown by red crosses in Fig.~\ref{fig:isolated}) as the correlation length in Fig.~\ref{figrc}.
The parameters are given in Appendix~\ref{sec:gaas}.
We compare four values of the nuclear spin temperature: $\Theta_\text{N}\pm =100$~$\mu$K and $\Theta_\text{N} =\pm 1$~$\mu$K.

The two spectra at  $\Theta_\text{N}\pm =100$~$\mu$K   are almost identical.
Indeed, when nuclear spin correlations are negligibly small, then the SN spectrum is just a Lorentzian function with the spectral width at half maximum inversely proportional to the electron spin relaxation time $\textrm{HWHM}=\Gamma_\text{S}/2\pi$.
The latter does not depend on the sign of the nuclear spin temperature in the absence of correlations, and is determined by the donor density (see Fig.~\ref{fig:tau}.

By contrast, at $\Theta_\text{N} =\pm 1$~$\mu$K, when nuclear spin system is cold, but still above the transition temperature, the correlations  build up.
They affect electron spin relaxation time in a way that depends on the sign of the nuclear spin temperature.
At  $\Theta_\text{N} >0 $ electron spin relaxation time shortens.
This is the consequence of the reduction of the correlation length in the vicinity of the polaron transition, making motional narrowing inefficient.
At  $\Theta_\text{N} <0 $ electron spin relaxation time grows up due to motional narrowing  that accompanies the increase of the correlation length.
Thus, the onset of correlations can be detected by measuring electron spin noise even above critical temperature.

\section{Perspectives for experimental observation of the nuclear magnetic ordering.}
\label{sec:towards}

The potential experimental detection of the electron-induced nuclear correlations and ordering rely on our ability to efficiently cool the NSS.
In order to be as realistic as possible, we focus on  n-GaAs, where both electron and nuclear spin dynamics have been extensively explored.

In n-doped GaAs ($n_\text{D}=2\times10^{15}$~cm$^{-3}$) nuclear spin temperatures as low as $\Theta_\text{N} \approx\-2$~$\mu$K have been reported at $T_\text{L}\approx4$~K.
This is encouraging, since this value is close to the critical temperature required to reach the ferromagnetic  order ($\Theta_\text{N}^\text{d}\approx -0.5$~$\mu$K).
The method usually employed for deep cooling of the nuclear spin consists of two steps:  (i) optical pumping that is mediated by the hyperfine interaction with spin-polarized electrons.
Under a magnetic field $B_i=200$~G the achieved nuclear spin polarization defines an initial temperature $\Theta_{\text{N}i}$, (ii) adiabatic demagnetization to zero field, that provides further cooling down to $\Theta_\text{N}=\Theta_{\text{N}i} B_\text{L}/B_i$.
The effective local field  $B_\text{L}$ determines the actual efficiency of the cooling.
It includes contributions from the dipole-dipole interaction ($B_\text{dd}\approx 2$~G) and the quadrupole interaction that can be induced by strain.

Keeping lattice temperature at $T_\text{L}\approx4-5$~K, the optimisation of pumping  efficiency, in particular using higher pumping field $B_i$ and reducing the strain in the sample, may be sufficient to reach negative  temperatures well below the critical value required for the formation of the nuclear ferromagnet in the sample with $n_\text{D}\approx 10^{15}$~cm$^{-3}$ (see Fig.~\ref{fig:isolated}).
Eventually, choosing the samples with lower donor densities may increase  the absolute value of the critical temperature and favor the formation of the nuclear ferromagnet, but we do not yet have enough data on the electron spin relaxation rates in such low-doped n-GaAs samples.
By contrast, most straightforward way of reaching positive critical temperature for the formation of nuclear polarons is to cool the crystal lattice in the sample with  $n_\text{D}=10^{16}$~cm$^{-3}$ (see Fig.~\ref{fig:isolated}) well below $T_\text{L}=2$~K.

\section{Conclusions}
\label{sec:conclusions}
 We have shown that in n-doped semiconductors with positive hyperfine constant, two kinds of magnetically ordered states can be induced by resident electrons in the deeply cooled nuclear spin system.
The magnetic phase diagram is determined by three parameters:  lattice temperature, donor density and the sign of the nuclear spin temperature $\Theta_\text{N}$.

When NSS is cooled down to a positive temperature below critical one $\Theta_\text{N}^\text{s}>0$, randomly oriented nuclear  spin polarons form under the orbit of each donor.
The underlying  mechanism relies  on the positive feedback, mediated by static polarization of nuclear and electron spins by
Knight and Overhauser fields respectively. The critical nuclear spin temperature for the formation of randomly oriented polarons state decreases when the lattice temperature is increased.
The models of nuclear polarons have been developed previously, but they neglected dynamic polarisation of the electron spin by the cold NSS.
We have shown that the formation of the nuclear polarons is not impeded by  dynamic polarisation even when hyperfine relaxation dominates over spin-orbit mechanism.
%
%Here we have shown that it is limited to $T_L$ lower than $k_\text{B} T_L=(1+\Gamma_L/\Gamma_\text{N})\langle A \rangle N I(I+1)/3$,  because at higher lattice temperatures the formation of the nuclear polarons is impeded by the dynamic polarisation of the electron spin by the NSS, that can not be neglected any more.
%

In NSS cooled down to a negative temperature below critical one $\Theta_\text{N}^\text{d}<0$, we predict the formation of an original long-range ordered state, that we call dynamically-induced nuclear ferromagnet.
It should manifest itself when electron spin dynamics is dominated by the hyperfine coupling, rather than by the spin-orbit interaction.
The underlying feedback mechanism can be understood as a dynamic polarization of the localized electron spin by the cold NSS polarized in the Knight field.
The dominance of the hyperfine coupling in low-doped systems and QDs is well known and confirmed by numerous experiments, but the positive feedback loop that leads in this case to the nuclear ferromagnetic state has been overseen soo far.

The lifetime of the ordered states is limited by the inevitable heating of the system, on the scale of the order of several seconds in n-GaAs.
Within this time, after cooling the NSS to a sufficiently low nuclear spin temperature, the nuclear spin ordering can be detected by different techniques: off-resonant Faraday rotation, spin noise spectroscopy, photoluminescence combined with radio-frequency absorption.

The strategy to reach magnetically ordered states may include lowering down the sample temperature down to and below $2$~K rather than $4-5$~K used in previous experiments, and  cooling the NSS at higher magnetic fields prior to adiabatic demagnetization.
Finally, samples with unstrained QDs may be promising.
Stronger electron localization as compared to donor-bound electrons in bulk n-GaAs ensures stronger interaction between electron and nuclear spins.
This may offer higher critical temperatures for both nuclear polarons and dynamically induced-induced nuclear ferromagnetism.

\begin{acknowledgments}
We wish to acknowledge the support of the joint grant  of the Russian Foundation for Basic Research (RFBR, Grant No. 16-52-150008)   and
 National Center for Scientific Research (CNRS, PRC SPINCOOL No. 148362). MSK and KVK acknowledge support from Saint-Petersburg State
University via a research Grant No. 73031758 and from RFBR Grant No 19-52-12043.
\end{acknowledgments}

%\appendix

\section{Appendix}
\setcounter{equation}{0}
\renewcommand{\theequation}{A\arabic{equation}}

\subsection{Derivation of the rate equations for the coupled electron-nuclei spin system }
\label{Sec:Eq1}
%Let us choose the direction of $z$-axis along average electron $\langle \vec{S} \rangle$ and nuclear $\langle \vec{J} \rangle$ spins in the coupled system.
%
The populations of electron spin  $S=1/2$ with projections $\pm  1/2$ on the $z$ axis chosen along the Overhauser field are equal to $S \pm \langle S_z \rangle$, respectively.
The rate equation for the average electron spin projection $\langle S_z \rangle$ reads:
\begin{equation}
\langle \dot{S}_{z} \rangle= -\langle {S}_{z} \rangle (p_++p_-)+S(p_+-p_-),
\label{eq:Sz}
\end{equation}
where $p_+$ and $p_-$ are probabilities of transitions with rising and lowering the electron spin projection by one, correspondingly. Such transitions occur with simultaneous change of states of the nuclear spin system and the lattice: the angular momentum is transferred to nuclei, while the energy goes to the lattice. In fact, these are transitions in the coupled electron-nuclear spin system, induced by interaction with the lattice. As shown in   [\onlinecite{Abragam}] (Ch 8), in the approximation of short correlation time the probabilities of such transitions with mutual electron-nuclear spin flips  can be written as:
\begin{align}
 p_{\downarrow m}^i=F_-\frac{\left(  A_i v_0 \Psi_i^2 \right)^2 \tau_\text{c}}{\hbar^2 }
 \lvert\langle (m_i+1,-S)\lvert\hat{S}_- \hat{I}_+\rvert (m_i,S) \rangle\rvert^2
 \nonumber \\
 p_{\uparrow m}^i =F_+\frac{\left(  A_i v_0 \Psi_i^2 \right)^2 \tau_\text{c}}{\hbar^2 }
 \lvert\langle (m_i-1,S)\lvert\hat{S}_+ \hat{I}_-\rvert (m_i,-S) \rangle\rvert^2.
\label{eq:pmarrow}
\end{align}
Here $\hat{I}_\pm$ are the rising and lowering nuclear spin operators,  $m_i$ is the spin projection of the i-th nuclear spin, $\Psi_i$ is the absolute value of the electron wave function at the i-th nuclei position, $v_0$ is the volume of the crystal elementary cell, $A_i$ is the hyperfine constant of the i-th nucleus and $F_{\pm}$ characterise the spectral power density of a random force describing interaction of the spin system with the lattice.
As follows from the principle of detailed balance,
$F_+/F_-=\exp(-\hbar \Omega_\text{N} \beta_\text{L} )$, where $\hbar \Omega_\text{N}$ is electron spin splitting in the Overhauser field created by the underlying nuclei.
Taking into account that
\begin{flalign}
\lvert \langle m_i-1 \lvert \hat{I}_-\rvert m_i \rangle \rvert^2=
 \langle m_i \lvert \hat{I}_+\hat{I}_-\rvert m_i \rangle =
 \langle m_i \lvert \hat{I}^2-\hat{I}_z^2+\hat{I}_z\rvert m_i \rangle
 \nonumber \\
 \lvert \langle m_i+1 \lvert \hat{I}_+\rvert m_i \rangle \rvert^2=
 \langle m_i \lvert \hat{I}_-\hat{I}_+\rvert m_i \rangle =
 \langle m_i \lvert \hat{I}^2-\hat{I}_z^2-\hat{I}_z\rvert m_i \rangle
  \nonumber \\
 \langle -S \lvert \hat{S}_+\rvert S \rangle=\langle S \lvert \hat{S}_-\rvert -S \rangle=1
\label{eq:calc}
\end{flalign}
and averaging  over all the projections of each nuclear spin,  $m_i$, with the distribution function $\rho_m$ corresponding to the spin temperature of the nuclear system we obtain the probabilities of the electron spin flip transitions due to interaction with i-th nucleus as
\begin{align}
 p_{-}^i =F_-\frac{\left(  A_i v_0 \Psi_i^2 \right)^2 \tau_\text{c}}{\hbar^2 }
\left[I(I+1)-\langle I_{iz}^2\rangle - \langle I_{iz}\rangle \right]
 \nonumber \\
 p_{+}^i =F_+\frac{\left(  A_i v_0 \Psi_i^2 \right)^2 \tau_\text{c}}{\hbar^2 }
\left[I(I+1)-\langle I_{iz}^2\rangle - \langle I_{iz}\rangle \right],
\label{eq:pmi}
\end{align}
where
\begin{equation}
\langle I_{iz}\rangle=\frac{\sum_{m_i=-I}^{I}
m_i\times \mathrm{exp}(-m_i \langle S_{z}\rangle A_i v_0 \Psi_i^2 \beta_\text{N})}
{\sum_{m_i=-I}^{I}\mathrm{exp}(-m_i \langle S_{z}\rangle A_i v_0 \Psi_i^2 \beta_\text{N})}
\label{eq:Iiz}
\end{equation}
is the average spin projection of the i-th nucleus on the z-axis (along Knight and Overhauser fields)  and
\begin{equation}
\langle I_{iz}^2\rangle=\frac{\sum_{m_i=-I}^{I}
m_i^2 \times \mathrm{exp}(-m_i \langle S_{z}\rangle A_i v_0 \Psi_i^2 \beta_\text{N})}
{\sum_{m_i=-I}^{I}\mathrm{exp}(-m_i \langle S_{z}\rangle A_i v_0 \Psi_i^2 \beta_\text{N})}
\label{eq:Iiz2}
\end{equation}
is the mean squared value of the same projection.
In order to obtain the full probabilities of flipping the electron spin up or down, one has to sum Eq. \ref{eq:pmarrow} over all the  nuclei situated under the orbit of a given electron:
\begin{align}
 p_{-} = (v_0 \langle A \rangle)^2 F_-\sum_{i}\frac{  \Psi_i^4 \tau_\text{c}}{\hbar^2 }
\left[I(I+1)-\langle I_{iz}^2\rangle - \langle I_{iz}\rangle \right]
 \nonumber \\
 p_{+} =(v_0 \langle A \rangle)^2  F_+\sum_{i}\frac{\Psi_i^4 \tau_\text{c}}{\hbar^2 }
\left[I(I+1)-\langle I_{iz}^2\rangle + \langle I_{iz}\rangle \right],
\label{eq:pmlast}
\end{align}
where $\langle A \rangle=\sum_l{\mathcal{A}_lA_{l}}$    is the hyperfine interaction constant averaged over all nuclear species in the crystal, $A_{l}$ and $\mathcal{A}_l$ are the hyperfine constant and the abundance of $l$-th isotope, respectively. We can now define electron spin relaxation rate due to hyperfine interaction as
\begin{equation}
\Gamma_\text{N}\equiv (v_0 \langle A \rangle)^2   (F_+ +F_-)\sum_{i}\frac{\Psi_i^4 \tau_\text{c}}{\hbar^2 }
\left[  I(I+1)-\langle I_{iz}^2\rangle  \right],
\label{eq:taunA}
\end{equation}
the thermal equilibrium value of the mean electron spin $S_\text{T}$:
\begin{equation}
2S_\text{T}=\frac{F_+-F_-}{F_+ +F_-},
\label{eq:ST}
\end{equation}
mean z-projection of the ensemble of the nuclear spins interacting with a given electron
\begin{equation}
\langle J_{z}\rangle=\frac{\sum_{i}  \Psi_i^4 \langle I_{iz}\rangle}
{\sum_{i} \Psi_i^4},
\label{eq:Iz}
\end{equation}
and mean squared transverse (perpendicular to z-axis) fluctuation of the ensemble of the nuclear spins interacting with a given electron.
\begin{equation}
\langle J_{\perp}^2\rangle=I(I+1)- \frac{\sum_{i} \Psi_i^4 \langle I_{iz}^2\rangle}
{\sum_{i} \Psi_i^4},
\label{eq:Iz2}
\end{equation}
By substituting  Eqs. (\ref{eq:pmlast}) into Eq. (\ref{eq:Sz}) and using the above definitions
we obtain:
\begin{equation}
\langle \dot{{S}_z} \rangle =-\Gamma_\text{N}\left( \langle S_z\rangle-S_\text{T}\right)+\Gamma_\text{N}\frac{\langle {J}_z \rangle }
{2\langle J_{\perp}^2\rangle}
\left ( 1- \frac{\langle {S}_z\rangle S_\text{T}}{S^2}\right).
\label{eq:StauN}
\end{equation}
If, in addition to relaxation by nuclei, there is some spin-orbit relaxation, this equation should be complimented by the term  $-\Gamma_\text{L}\left( \langle S_z\rangle-S_\text{T}\right)$ on the right-hand side. In this case the full equation describing both hyperfine and spin-orbit  relaxation in the ensemble of localized electrons takes the  form given by  Eq. (\ref{eq:rate}) in the main test.
%\begin{equation}
%\langle \dot{{S}_z} \rangle =-\Gamma_S
%\left( \langle S_z\rangle-S_T\right)+\Gamma_N\frac{\langle {I}_z \rangle }
%{2\langle I_{\perp}^2\rangle}
%\left ( 1- \frac{\langle {S}_z\rangle S_T}{S^2}\right).
%\label{eq:StauS}
%\end{equation}
%

As  far  as  collective  electron-nuclei  spin  states  are not formed, both electron and nuclear spin polarisation remain weak. Under these conditions $\langle J_{\perp}^2\rangle \approx 2I(I+1)/3$ and $\langle S_{z}\rangle S_\text{T} \rightarrow 0$. Assuming the exponential form of the electron wavefunction $\Psi_i \propto \exp(-r_i/a_\text{B})$, where $a_\text{B}$ is the Bohr radius of the donor-bound electron we can calculate  $\langle J_{z}\rangle \approx -I (I+1) \beta_\text{N} \langle A\rangle \langle S_z \rangle/3N$. Here $N$ is defined as
\begin{equation}
\frac{1}{N}\equiv v_0  \frac{ \sum_{i} \Psi_i^6 }{\sum_{i} \Psi_i^4 }=
\frac{8 v_0}{27 \pi a_\text{B}^3}.
\label{eq:N}
\end{equation}
It can be considered as a number of nuclei under the orbit of donor-bound electron, for shallow donors in GaAs $N \approx 2.4\times 10^5$. Thus, Eq.~(\ref{eq:rate}) reduces to the linear differential equation:
\begin{equation}
\langle \dot{{S}_z} \rangle =-\Gamma_\text{S}
\left( \langle S_z\rangle-S_\text{T}\right)+\Gamma_\text{N} \frac{\langle {J}_z \rangle}
{Q},
\label{eq:Slined}
\end{equation}
with $Q\equiv 4I(I+1)/3$ and
\begin{equation}
S_\text{T} =-\frac{1}{2}\textrm{tanh} \left(\frac{\hbar \Omega_\text{N} \beta_\text{L}}{2}\right)
\approx -\frac{\langle A\rangle \langle J_z\rangle \kappa \beta_\text{L} }{4},
\label{eq:STappendix}
\end{equation}
where $\Omega_\text{N}$ is  the  angular  frequency  of electron  spin  precession  in  root mean square fluctuation of the  Overhauser field
\begin{equation}
\hbar\Omega_\text{N}=\sqrt{Q \kappa \langle A \rangle ^2  /4N},
\label{eq:hbar2Omega2}
\end{equation}
$\kappa=27 n_0/64$ and $n_0$ is the number of atoms in the elementary cell of the crystal. Note, that the factor $\kappa$ appears in Eq.~(\ref{eq:STappendix}) due to the choice that we have made in the definition of $N$ (cf Eq.~(\ref{eq:N})).

\subsection{Parameters of the coupled electron-nuclear spin system in n-GaAs: interaction, diffusion and relaxation}
\label{sec:gaas}

\begin{figure}
  \centering
  \includegraphics[width=8.5 cm]{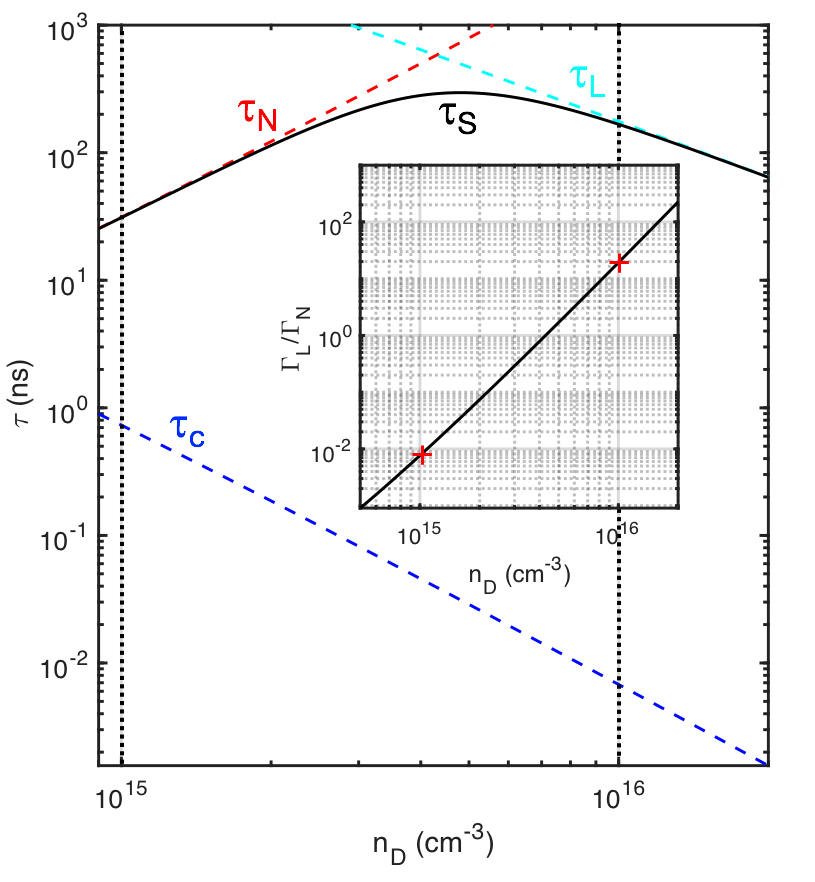}
  \caption{Characteristic times for the electron spin in the insulating n-GaAs as a function of the donor density (dashed lines): correlation time (blue), hyperfine relaxation time (red), spin-orbit relaxation time (cyan).  Inset shows the ratio of two relaxation rates as a function of donor density. Vertical dotted lines and red crosses (inset) point the parameters corresponding to Figs.~\ref{figrc} and ~\ref{fig:spectra}.
  %One can see that carrier-induced ferromagnetism is favored at low densities, while polaron states are expected at higher densities.
  }
  \label{fig:tau}
\end{figure}
Electron spin relaxation has been exhaustively studied in the insulating n-GaAs.
Correlation time of the electron spin was measured over a broad range of donor concentrations $n_\text{D}$ \cite{Dzhioev2002,Belykh2017}.
Its dependence on $n_\text{D}$ can be approximated by the expression:
\begin{equation}
\tau_\text{c}(n_\text{D})=0.2\left( \frac{n_\text{D}}{10^{15}} \right )^{2.3} \text{log}\left(  15\frac{n_\text{D}}{10^{15}} \right ),
\label{eq:tauc}
\end{equation}
where $n_\text{D}$ is expressed in inverse cubic centimetres  and $\tau_\text{c}$ in nanoseconds.

Electron spin relaxation rate due to hyperfine coupling is given by
\begin{equation}
\Gamma_\text{N}= \frac{2}{3} \Omega_\text{N}^{2} \tau_\text{c},
\label{eq:tausn}
\end{equation}
where $\Omega_\text{N}$ is the angular frequency that characterizes electron spin precession in the
fluctuating Overhauser field  defined in the previous section, see Eq.~\ref{eq:hbar2Omega2}.
The spin-orbit relaxation rate is also related to the correlation time and donor density:
\begin{equation}
\Gamma_\text{L}= \frac{n_\text{D}^{-2/3}}{L_\text{SO}^2\tau_\text{c} },
\label{eq:tausa}
\end{equation}
where $L_\text{SO}$ is so-called spin-orbit length \cite{KavokinSST09}. In GaAs $L_\text{SO}\approx 7$~$\mu$m.
Finally, the electron spin diffusion constant, determined by electron hopping and exchange interaction in the impurity band, reads:
\begin{equation}
D_\text{s}= \frac{n_\text{D}^{-2/3}}{3\tau_\text{c} },
\label{eq:Ds}
\end{equation}
Fig.~\ref{fig:tau} shows low temperature ($T_\text{L}<5$~K) correlation time, as well as two relevant electron spin relaxation times calculated according to Eqs.~(\ref{eq:tauc}-\ref{eq:tausa}) as a function of the donor density, while inset shows the ratio $\Gamma_\text{L}/\Gamma_\text{N}$.
%
%This one-to-one relation between donor density and the ratio $\Gamma_L/\Gamma_N$
The right scale in Fig.~\ref{fig:isolated} shows the donor densities  calculated using Eqs.~(\ref{eq:tauc}-\ref{eq:tausa}).

Table~\ref{tab:param2} summarises the values of spin, gyromagnetic ratio, hyperfine constants and abundance for each of the three isotopes in GaAs.
Other parameters used in the calculations are listed in Table~\ref{tab:param}.
%One can see that in GaAs doped with $n_\text{D}=10^{15}$~cm$^{-3}$ the critical temperature for the electron-induced nuclear ferromagnet can be as high as $\Theta_N^c=0.3$~$\mu$K at $T_L=5$~K.

%\subsection{Parameters of the nuclear spin system in GaAs}
%\label{sec:app:param}
%Table~\ref{tab:param} summarises the values of spin, gyromagnetic ratio, hyperfine constants and abundance for each of the three isotopes in GaAs.
%
%Other parameters used in the calculations are listed in Table~\ref{tab:param2}.
%
\begin{table}[h!]%{ |s|p{3cm}|p{3cm}|  }
\begin{tabular}{|l|c|}
\hline
%\rowcolor{lightgray} \multicolumn{3}{|c|}{Country List} \\
parameter &  value \\
\hline
\hline
Donor Bohr radius, $a_\text{B}$ &  $10$~nm \\
\hline
Volume of the elementary cell, $v_0$     & $4.5\times10^9$~m$^{-3}$  \\
\hline
Atoms number in the elementary cell, $n_0$     & $2$  \\
\hline
Spin-orbit length, $L_\text{SO}$     & $7.5$~$\mu$m  \\
\hline
Electron gyromagnetic ratio, $\gamma_{e}$     & $0.64$~MHz/G  \\
\hline
\end{tabular}
\caption{Parameters used in numerical calculations for n-GaAs }
\label{tab:param2}
\end{table}
\begin{table}[h!]%{ |s|p{3cm}|p{3cm}|  }
\begin{tabular}{|l|c|c|c|}
\hline
%\vspace{2}
%\rowcolor{lightgray} \multicolumn{3}{|c|}{Country List} \\
isotope  &  $^{75}As$ &  $^{71}Ga$ &$^{69}Ga$ \\
\hline
\hline
Spin, $I_l$&  3/2 &  3/2  &  3/2 \\
\hline
Abundance, $\mathcal{A}_l$&  0.5 &  0.2  &  0.3 \\
\hline
Hyperfine constant, $A_l$ ($\mu$eV)&  43.5 &  54.8  &  43.1 \\
\hline
Gyromagnetic ratio, $\gamma_{N,l}$ ($10^7$rad/Ts)&  4.6 &  8.1  &  6.44 \\
\hline
\end{tabular}
\caption{Values of spin, gyromagnetic ratio, hyperfine constants and abundance of each of three isotopes in GaAs \cite{Chekhovich2017,Harris2002}. }
\label{tab:param}
\end{table}

\subsection{Calculation of electron spin noise in the presence of nuclear spin correlations }
\label{Sec:SNS}
In order to calculate the spectral power density of electron spin fluctuations in the regime where the fluctuations of  nuclear spin can be correlated, we need to develop a method based on $k$-components of the nuclear spin fluctuations.
Let us consider a cubic box with the volume $V\gg n_\text{D}^{-1}$.
Electron and nuclear spin densities in the box can be expanded in the Fourier series with $k_{\sigma,n}=2\pi n/ V^{1/3}$, where $\sigma 	\in \{x,y,z\}$ and $0<n<(V n_\text{D})^{1/3}$.
The total number of $k$-modes $Vn_\text{D}$ is equal to the number of donors in the volume.

The zero-$k$ mode of the $z$-component of the electron spin density under periodic pump $S_Ge^{—i\omega t}$ can be written as
\begin{equation}
\dot{S}_{z,0}= \sum_k{\Omega_{x,\vec{k}} S_{y,-\vec{k}}}-
\sum_k{\Omega_{y,\vec{k}} S_{x,-\vec{k}}}+S_Ge^{i\omega t}
\label{eq:Sz0}
\end{equation}
where $\Omega_{x,\vec{k}}$ and  $\Omega_{x,\vec{k}}$ are Fourier components of the nuclear fluctuation field in frequency units.
Since the spatial harmonics of $x$ and $y$-components of the electron spin are much smaller than the $z$-component, we keep only the terms containing  $S_{z,0}$ in the corresponding equations:
\begin{align}
\dot{S}_{x,-\vec{k}}=
\Omega_{y,-\vec{k}}{S}_{z,0}-D_\text{s} k^2 S_{x,-\vec{k}} \nonumber \\
\dot{S}_{y,-\vec{k}}=
-\Omega_{x,-\vec{k}}{S}_{z,0}-D_\text{s} k^2 S_{y,-\vec{k}}
\label{eq:Sk}
\end{align}
These equations have the following solutions:
\begin{align}
{S}_{x,-\vec{k}}(t)=
\Omega_{y,-\vec{k}}  \int_{0}^{\infty} e^{-D_\text{s} k^2 \tau}
{S}_{z,0}(t-\tau)\,d\tau \nonumber \\
{S}_{y,-\vec{k}}(t)=
-\Omega_{x,-\vec{k}}  \int_{0}^{\infty} e^{-D_\text{s} k^2 \tau} {S}_{z,0}(t-\tau)\,d\tau
\label{eq:Sk_sol}
\end{align}
Substituting Eqs.~(\ref{eq:Sk_sol}) into Eq.~(\ref{eq:Sz0}) we obtain
\begin{equation}
\dot{S}_{z,0}= -\frac{2}{3}\sum_k \Omega^2_{N,\vec{k}}
\int_{0}^{\infty} e^{-D_\text{s} k^2 \tau}
{S}_{z,0}(t-\tau)\,d\tau +S_G e^{i\omega t},
\label{eq:Sz0Sol}
\end{equation}
where we assumed that nuclear spin fluctuations are isotropic $\frac{2}{3}\Omega^2_{N,\vec{k}}=\Omega_{x,\vec{k}}+\Omega_{y,\vec{k}}$.
The solution of Eq.~(\ref{eq:Sz0Sol}) has the form $S_{z,0}(t)=S_z(\omega)e^{—i\omega t}$, and we come to the equation for  $S_z(\omega)$:
\begin{equation}
i \omega {S}_{z}(\omega)= -\Phi(\omega){S}_{z}(\omega)+S_G
\label{eq:Sz0mega}
\end{equation}
with
\begin{equation}
\Phi(\omega)= -\frac{2}{3}\sum_k \frac{\Omega^2_{N,\vec{k}}}{D_\text{s} k^2+\omega^2}
\label{eq:phi}
\end{equation}

Replacing in Eq.~(\ref{eq:phi}) the summation by the integration over $k$-space, we get
\begin{equation}
\Phi(\omega)=\frac{V}{3 \pi^2}
\int_{0}^{\alpha n_\text{D}^{1/3}} \frac{k^2 \Omega_{N,\vec{k}}^2(D_\text{s} k^2-i \omega)}{(D_\text{s} k^2)^2+\omega^2} dk,
\label{eq:phiInt}
\end{equation}
where the upper integration limit should be determined from the conditions in the absence of correlations:  $\Phi(\omega=0)=\Gamma_\text{S}$ and  $\Omega^2_{N,\vec{k}}(\omega=0)=\Omega_\text{N}^2/V n_\text{D}\equiv \Omega^2_{0,\vec{k}}$.
Recalling that $D_\text{s}$ is given by Eq.~(\ref{eq:Ds}), we get
\begin{equation}
\Phi(\omega)=\frac{\Omega_\text{N}^2}{3 \pi^2 n_\text{D}}
\int_{0}^{\nu n_\text{D}^{1/3}} \frac{k^2 (\Omega_{N,\vec{k}}^2/\Omega_{0,\vec{k}}^2)(D_\text{s} k^2-i \omega)}{(D_\text{s} k^2)^2+\omega^2} dk,
\label{eq:phiInt2}
\end{equation}
where $\nu=2/3 (\pi^2 \Gamma_\text{S}  / \Gamma_\text{N})$ and the ratio $\Omega_{N,\vec{k}}^2/\Omega_{0,\vec{k}}^2$ is nothing but nuclear spin fluctuation power given by Eq.~(\ref{eq:fluct}), normalized by its maximum value:
\begin{equation}
\Omega_{N,\vec{k}}^2/\Omega_{0,\vec{k}}^2=3 \langle J_z^2\rangle_{0,\vec{k}}/I(I+1).
\label{eq:OmJ2}
\end{equation}

The  solution of Eq.~(\ref{eq:Sz0mega}) reads:
\begin{equation}
{S}_{z}(\omega)=\frac{S_G}{i \omega +\Phi(\omega)) }
\label{eq:SzSol}
\end{equation}

Now, considering $S_G$  as a time harmonic of a  $\delta$-correlated random Langevin force, we find the expression for the spectral power density of the electron spin fluctuations:
\begin{multline}
\langle S^2_{z}\rangle _{\omega}=\langle S_{z}(\omega)S_{z}(-\omega)\rangle = \\
\frac{{S_G}^2}{\left (\text{Re}(\Phi(\omega))\right )^2+\left (\omega+\text{Im}(\Phi(\omega))\right )^2},
\label{eq:S2}
\end{multline}
where $\Phi(\omega)$ is given by Eqs.~(\ref{eq:phiInt2}-\ref{eq:OmJ2}) and $ \langle J_z^2\rangle_{0,\vec{k}}$ by Eq.~(\ref{eq:fluct}).

% The \nocite command causes all entries in a bibliography to be printed out
% whether or not they are actually referenced in the text. This is appropriate
% for the sample file to show the different styles of references, but authors
% most likely will not want to use it.
%\nocite{*}

\bibliography{NuclearSpinLibrary}% Produces the bibliography via BibTeX.

\end{document}